\documentclass[10pt,superscriptaddress,onecolumn,showpacs,showkeys,aps,prl,preprint]{revtex4-1}
\usepackage{graphicx}
\usepackage{dcolumn}
\usepackage{subfigure}
\usepackage{bm}
\usepackage{ifpdf,braket}
\usepackage{epstopdf}
\ifpdf
\usepackage[pdftex,
        colorlinks=true,
        pdftitle={Coexistence of magnetic and charge order in a two-component order parameter description of the layered superconductors},
        pdfauthor={Mauro M. Doria},
        pdfsubject={Vortex theory},
        pdfkeywords={Departamento de Fisica, Universidade Federal do Rio de Janeiro, Brazil},
        baseurl={http://www.if.ufrj.br},
        pdftoolbar=true,
        pdfmenubar=false,
        pdfpagemode=UseOutlines,
        pdfstartview=FitH,
        bookmarksopen=false
        ]{hyperref}
\fi

\begin{document}

\title{Coexistence of magnetic and charge order in a two-component order parameter description of the layered superconductors}
\author{Mauro M. Doria}%
\affiliation{Departamento de F\'{\i}sica dos S\'{o}lidos, Universidade Federal do Rio de Janeiro, 21941-972 Rio de Janeiro, Brazil}%
\email{mmd@if.ufrj.br}
\author{Alfredo A. Vargas-Paredes}%
\affiliation{Departamento de F\'{\i}sica dos S\'{o}lidos,
Universidade Federal do Rio de Janeiro, 21941-972 Rio de Janeiro,
Brazil}
\author{Marco Cariglia}%
\affiliation{Departamento de F\'{\i}sica , Universidade Federal de
Ouro Preto, 35400-000 Ouro Preto Minas Gerais, Brazil}

\date{\today}

\begin{abstract}
A two-component order parameter approach for the layered superconductor is shown to form a condensate with magnetic and charge degrees of freedom. This condensate is an inhomogeneous state, topologically stable, that exists without the presence of an applied magnetic field. We show that the charge density in the layers presents hexadecapole moment in its lowest order. Our approach is based on the first order equations that we show here to solve the variational equations for the special temperature defined by the crossing of the superconducting dome and the pseudogap transition line. Time reversal symmetry is broken and the weak local magnetic field produced by this inhomogeneous state falls below the threshold of experimental observation. We find that the charge distribution in the layers has an  hexadecapole moment in its lowest order.
\end{abstract}


\maketitle

\textbf{Introduction}. -- The concept of an order parameter was introduced in 1937 by Lev Landau to describe the second order phase transition in the specific heat of tin that takes place at the passage to the superconducting state and had been observed a few years before. In 1950 the celebrated Ginzburg-Landau (GL) theory was proposed to provide a gauge invariant macroscopic description of superconductivity and incorporated the second order phase transition and also London's theory which accounted for the Meissner effect. Interestingly this macroscopic description was developed without any knowledge of the microscopic mechanism of superconductivity, such as the existence of pairing, which was only proposed in 1956 by Leon Cooper. According to the BCS theory of superconductivity paired electrons condense into a single state whose description is attainable through the order parameter approach near to the critical temperature, $T_c$.  The discovery of the high-T$_c$ superconductors~\cite{bedmull86} brought a renewed interest in the order parameter approach because of the lack of understanding of the pairing mechanism. Hereafter high-$T_c$ superconductors are those that display a layered structure such that superconductivity originates in the layers. We argue in this paper that the description of the high-$T_c$ superconductors demands two complex order parameters. Multiple order parameter theories provide the optimal framework for the description of multiple phases and also of inhomogeneous condensates~\cite{sigrist91}.

Soon after the discovery of the high-$T_c$ superconductors by Bednorz and M\"uller~\cite{bedmull86} in 1987 two types of order parameter approaches were applied to them~\cite{brandt95}, namely, the anisotropic Ginzburg-Landau and the Lawrence-Doniach theories. The former is just the traditional GL theory with a mass anisotropy tensor to cope with the inertia acquired by Cooper pairs to move perpendicularly to the layers. This theory does not have layers and therefore cannot recognize them as the sources of the
superconducting state. In the latter theory superconductivity  only exists within the layers and the space between them is a perfect void. There is coupling between nearest neighbor layers through the Josephson effect. As successful applications of these two theories to the high-$T_c$ superconductors we quote the description of the torque~\cite{bosma11} and of the THz spectroscopy~\cite{kadowaki13}, respectively. Here we consider a third kind of order parameter approach  that takes the layers as the source of superconductivity and yet has the condensate outside them existing in an evanescent way. The high-$T_c$ superconductor is a stack of layers embedded in a metallic media since the condensate still exists in the inter-layer space, although it decays exponentially away from the layers~\cite{cariglia14}. The layers contain supercurrent circulation which demand distinct order parameter behavior just above and  below them. In this paper we show that these distinct properties lead to an inhomogeneous condensate with intrinsic magnetic and charge orders. The intertwinement of pairing, charge, and spin degrees of freedom has been the subject of intense research lately~\cite{berg09,vojta09}.

In the past decade it became clear that superconductivity exists above $T_c$, fact that can only be handled by the anisotropic GL and the Lawrence-Doniach theories from the point of view of thermal fluctuations of the order parameter. However recent understanding of the so-called temperature versus doping phase diagram shows that superconductivity above $T_c$ cannot be simply explained by thermal fluctuations.
The high-$T_c$ superconductor acquires new properties according to
the doping, namely, the number of carriers available for conduction
in the layers. The $T_c$ versus doping line of this diagram defines
a dome shaped curve. The superconducting state is called underdoped, optimally doped, and overdoped, respectively, according to the doping level relative to the maximum $T_c$. Thus besides the superconducting
state there are other states~\cite{khasanov07}, such as the so-called pseudogap state~\cite{alloul89}, that are not caused by thermal fluctuations. The pseudogap emerges at a temperature T$^*$, claimed to be a phase transition line by some~\cite{he11}. In the underdoped regime this temperature is above T$_c$ and decreases with increasing doping level. At some doping T$^*$=T$_c$, and beyond, one expects that the pseudogap line enters the superconducting dome to finally reach a
quantum critical point at $T=0$~\cite{daou10}. The
microscopic nature of the pseudogap remains controversial. In
this paper we assume that the pseudogap is also a condensate, and so can be described by the order parameter approach. The presence of two transition lines, namely, T$_c$ and T$^*$ is suggestive of a two-component order parameter $\Psi$, while the original anisotropic GL and
Lawrence-Doniach theories have only one, $\psi$. Multi-component
order parameter theories have been proposed for the high-T$_c$
superconductors since long ago~\cite{volovik85}.

\begin{figure}
\includegraphics[width=\columnwidth]{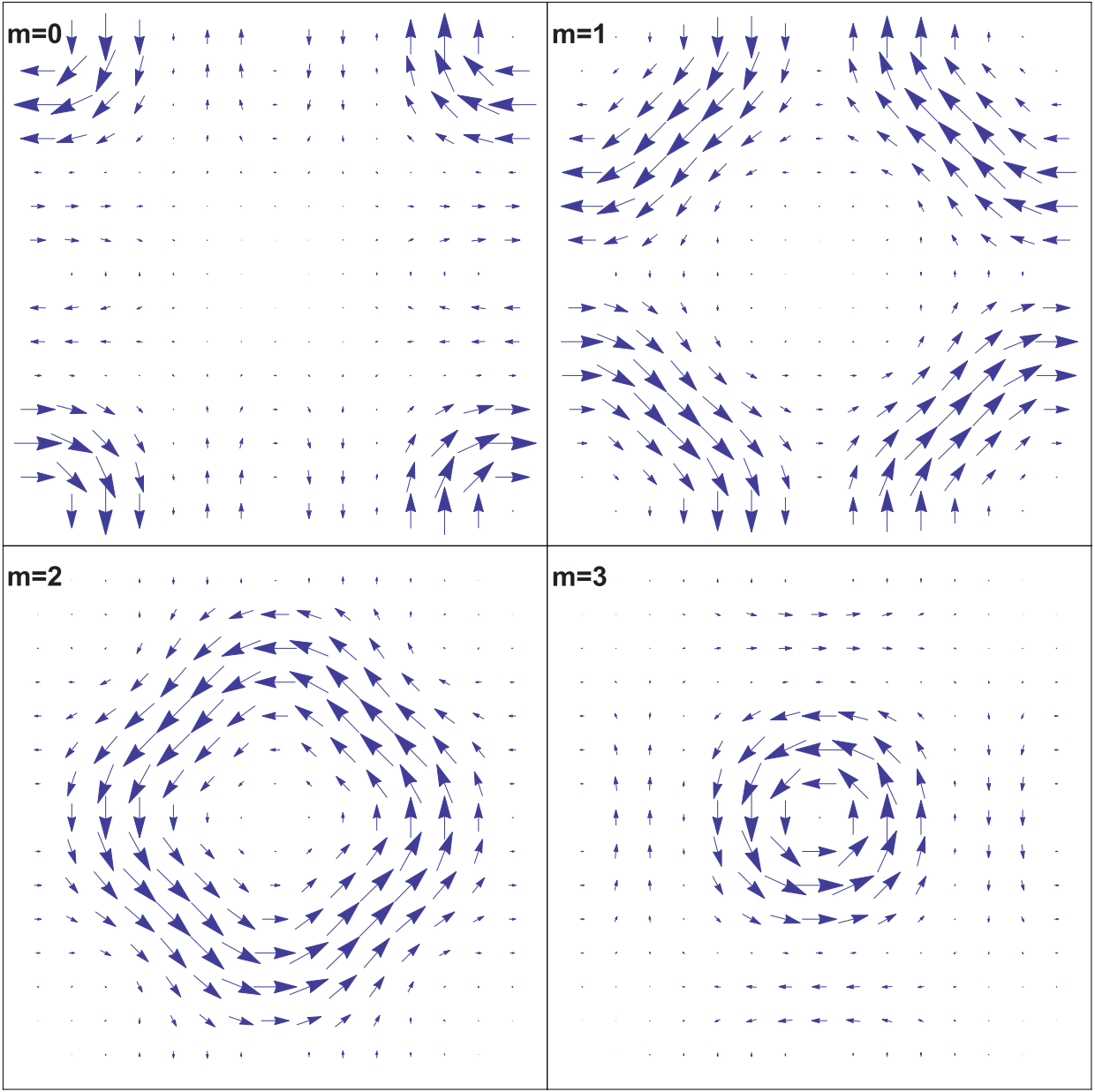}
\caption{The  superficial supercurrent, $J_s$, given by Eq.(\ref{currents}), is shown within the unit cell area, defined by $0\le x_i/L\le 1$, $i=1,2$, for each of the $m=0,1,2,\mbox{and}\;3$ states. These states are eigenvectors of $J_3$ defined by Eq.(\ref{j3e}). Notice that the superficial charge only exists in the layers and not in the interlayer space.}
\label{mstates}
\end{figure}
\begin{figure}
\includegraphics[width=\columnwidth]{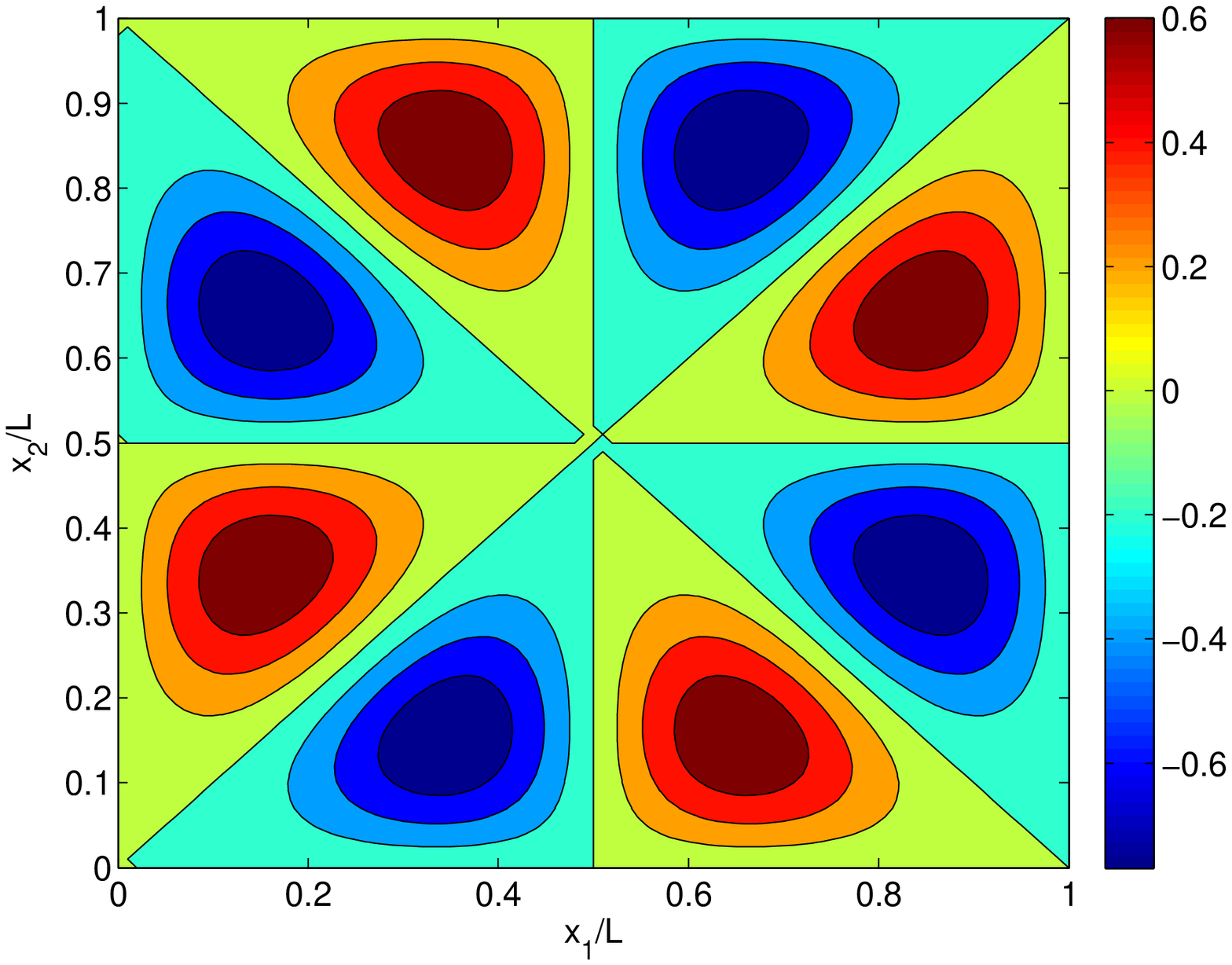}
\caption{The $\Sigma(x_1,x_2)$ function, defined in Eq.(\ref{sigma1}) is shown within the unit cell area, defined by $0\le x_i/L\le 1$, $i=1,2$. The charge density rate of all the $m$ states, shown in Fig.(\ref{mstates}), is proportional to $\Sigma(x_1,x_2)$. Notice the presence of positively and negatively charged spots within the unit cell rendering a hexadecapole moment.}
\label{charge}
\end{figure}

\textbf{The topological equations}. --
Interestingly, the prediction of a crystalline ordered state
made of topological excitations, i.e., the vortex lattice, was done based on  the so-called first order equations (FOE), and not on the second order variational GL equations. This important remarks stems directly from A. A. Abrikosov's original work~\cite{abrikosov57}, where the GL free energy only enters to determine which vortex lattices, among the possible ones, has the lowest energy. The FOE were rediscovered by E. Bogomolny~\cite{bogomolny76} in the context of string theory and shown to solve exactly  the GL second order equations for a particular value of the coupling constant ($\kappa=1/\sqrt{2}$).
The Abrikosov-Bogomolny equations are given by,
$D_{+}\psi=0$ and $h_3=C_3-(hq/mc)|\psi|^2$, where $C_3$ is a
constant and $D_{\pm} \equiv D_1\pm i D_2$. The covariant derivative, $D_i$, $i=1,2,3$, is described below. These equations demand
uniaxial symmetry, chosen  along the external magnetic field
direction, $\vec H = C_3 \hat x_3$. Then the single component order
parameter and the local magnetic field must be given by
$\psi(x_1,x_2)$ and $\vec h = h_3(x_1,x_2)\hat x_3$, respectively.
In case of no external field these equations give the trivial solution of a spatially homogeneous state. The first Abrikosov-Bogomolny equation
becomes $\nabla_{+}\psi=0$, and can be expressed as
$\partial\psi(z,z^{*})/\partial z^{*}=0$, $z=x_1+ix_2$. The only
possible solution, assuming periodicity in the plane, is of a spatially
constant order parameter, according to Liouville's theorem. By
selecting the constant $C_3=(hq/mc)|\psi|^2$ in the second
Abrikosov-Bogomolny equation then $h_3=0$.
Interestingly there is another set of FOE, the Seiberg-Witten equations, which  describe four dimensional massless magnetic monopoles~\cite{seiberg94}. Thus the FOEs form a family that render  topological solutions and for this simple reason we call them the topological equations. We claim that another pair of such equations is required to describe the topological excitations of the high-$T_c$ superconductors~\cite{doria10}:
\begin{eqnarray}
&&\vec{\sigma}\cdot\vec{D}\, \Psi=0, \label{foe1} \quad  \Psi =
\left(
\begin{array}{c} \psi_u \\ \psi_d \end{array} \right),\quad \mbox{and} \\
&&\vec h = \vec H - \frac{h q}{m c}\Psi^\dag  \vec{\sigma}\, \Psi.
\label{foe2}
\end{eqnarray}
$\vec D = (\hbar/i)\vec \nabla -
(q/c)\vec A$ is the covariant derivative, the local magnetic field is $\vec h = \vec \nabla \times \vec A$, $\vec \sigma$ are the Pauli matrices, and $\vec H$ is the applied external field. The FOE are not a consequence of the second order equations that follow from a variational principle applied to the free energy $F$. Nevertheless they solve them in some given approximation for no applied field leading to an inhomogeneous state, as shown in this paper.

Next we point that the topological Eqs.(\ref{foe1}) and (\ref{foe2}) naturally determine a local magnetic field since their solution  automatically satisfies Maxwell's equations, namely, Amp\`ere's law and also $\vec \nabla \cdot \vec h =0$ because,
\begin{eqnarray}
\vec \nabla \cdot \left (\Psi^\dag \vec{\sigma}\Psi \right)=0.
\end{eqnarray}
To proof this relation just check that Eq.(\ref{foe1}) can be expressed as $\vec \sigma \cdot \vec \nabla
\Psi = 2\pi i \vec \sigma \cdot \vec A \Psi/\Phi_0$, where
$\Phi_0=hc/q$ is the flux unit. Using that
$\nabla_i\left(\Psi^{\dag}\sigma_i\Psi \right)=\left (
\sigma_i\nabla_i\Psi\right)^{\dag}\Psi+\Psi^{\dag}\left (
\sigma_i\nabla_i\Psi\right)$, it follows that,
$\nabla_i\left(\Psi^{\dag}\sigma_i\Psi \right)=\left(2\pi i \vec
\sigma \cdot \vec A \Psi\right)^{\dag}\Psi/\Phi_0+\Psi^{\dag}\left
(2\pi i \vec \sigma \cdot \vec A \Psi\right)/\Phi_0=0$.

We seek the solution of Eqs.(\ref{foe1}) and (\ref{foe2}) for a stack of layers separated by distance $d$ and without the presence of an applied field ($\vec H = 0$). We shall show that $\Psi$ arises in the layers and evanesces away from them such as in a metallic medium able to sustain a three-dimensional state. For the moment assume an inhomogeneous solution of the FOEs, $\Psi \equiv \Psi(\vec x)$ and $\vec h \equiv \vec h(\vec x)$. Then the local inhomogeneous field implies on a spatially circulating supercurrent, both volumetric and superficial, $\vec J_s(\Psi)$, or, equally, a superficial magnetization, $\vec M_s(\Psi)= - c\, \hat x_3 \times \vec J_s(\Psi)$, where axis 3 is perpendicular to the layers. Thus it results from Eqs.(\ref{foe1}) and (\ref{foe2}) that,
\begin{eqnarray}\label{currents}
&& \vec J=- c\mu_B\vec \nabla \times \left(\Psi^\dag \vec{\sigma}\Psi\right) \\
&& \vec J_s = - 2c\mu_B \, \hat x_3 \times \Psi^\dag (0^+)
\vec{\sigma}\Psi(0^+),
\end{eqnarray}
where $\mu_B=\hbar q/2mc$ is the Bohr's magneton.
The presence of a superficial supercurrent signals a discontinuity of the field parallel to the layer, $\vec h_{\parallel} \equiv h_1\hat x_1 + h_2\hat x_2$, across the layers. To understand this discontinuity take the layer at $x_3=0$, for instance. The parallel field satisfies $\hat x_3
\times [\vec h\left(0^+\right)-\vec h\left(0^-\right)] = 4\pi \vec
J_s/c$, while the perpendicular field, $h_3$, must be continuous, $ \hat
x_3 \cdot [\vec h\left(0^+\right)-\vec h\left(0^-\right)] = 0$.
As a consequence of Maxwell's equations the volumetric and the superficial supercurrents are not truly independent, but related to each other to warrant that the supercurrent remains divergenceless:
\begin{eqnarray}\label{div-layer}
\vec \nabla \cdot \vec J_{s} + \hat x_3 \cdot \vec J\left(0^{+}\right)- \hat x_3 \cdot \vec J\left(0^{-}\right)=0
\end{eqnarray}
Hence the above equation is just a consequence of Amp\`ere's law applied to  a stack of layers with metallic medium in between them. To prove the above relation just apply the divergence operator to the parallel boundary condition, which gives that, $\vec \nabla \cdot \left[\hat x_3 \times \vec
h\left(0^+\right)\right]-\vec \nabla \cdot \left [\hat x_3
\times\vec h\left(0^-\right) \right] = 4\pi \vec \nabla \cdot\vec
J_s/c$. Then from  Amp\`ere's law it follows that $\vec \nabla \cdot
\left(\hat x_3 \times \vec h \right)=-4\pi \hat x_3\cdot \vec J/c$,
which leads to Eq.(\ref{div-layer}).

Therefore according to the present model the volumetric supercurrent is constantly entering and exiting each given layer and transforming itself into the superficial supercurrent. This means that there is charge entering and exiting the layer at a constant rate.  Eq.(\ref{div-layer}) describes
the net volumetric supercurrent between the layers that transforms
itself into the superficial supercurrent at each spatial point where $\hat x_3 \cdot \vec J\left(0^{+}\right)- \hat x_3 \cdot \vec J\left(0^{-}\right)
\ne 0$. We interpret this as a surface charge density within the layer, $\sigma$, defined by,
\begin{eqnarray}\label{sigma}
&&\vec \nabla \cdot \vec J_{s} + \frac{\partial \sigma}{\partial t}  = 0, \quad \mbox{where}\nonumber \\
&& \frac{\partial \sigma}{\partial t} \equiv  \hat x_3 \cdot \vec J\left(0^{+}\right)- \hat x_3 \cdot \vec J\left(0^{-}\right).
\end{eqnarray}
In summary the present model determines the rate of charge density, $\partial \sigma / \partial t$, that enters and exists at each point of a given layer. We find remarkable that a magnetostatic description of a stack of two-dimensional layers embedded in a metallic medium leads to an inhomogeneous charge rate density within a layer~\cite{edinardo14}.

\textbf{Time reversal symmetry and the topological charge}. --
Interestingly the FOEs, given by Eqs.(\ref{foe1}) and (\ref{foe2}), automatically break time reversal symmetry as they admit two solutions, one associated to a local magnetic field and the other to the reverted field.
For simplicity consider the no applied field case ($\vec H=0$) in Eqs.(\ref{foe1}) and (\ref{foe2}).
Assume $\Psi$ to be a known solution and consider another state $\Psi'=U\Psi^{*}$, where
$U=e^{i\alpha}\sigma_2$ ($UU^{\dagger}=1$) where $\alpha$ can be any angle. Then one obtains that $U\vec \sigma^{*}U^{\dagger}=-\vec \sigma$.
The time reversal operation flips the magnetic field and also the spin and it is well known that this can be achieved by a unitary rotation proportional to $\sigma_2$, the only imaginary Pauli matrix. The conjugation operation over the real expectation value of the spin gives that, $\left (\Psi^{\dagger}\vec \sigma \Psi \right)^*=\left(\Psi^{*}\right)^{\dagger}\vec {\sigma}^{*}\Psi^{*}=\left(\Psi^{*}\right)^{\dagger}U^{\dagger}U\vec {\sigma}^{*}U^{\dagger}\left(U\Psi^{*}\right)=-\left(U\Psi^{*}\right)^{\dagger}\vec {\sigma}\left(U\Psi^{*}\right)$. Therefore we have shown that $\Psi'^{\dagger}\vec {\sigma}\Psi'=-\Psi^{\dagger}\vec{\sigma}\Psi$, which according to Eq.(\ref{foe2}), also implies that $\vec{h}\left( \Psi'\right)= -\vec{h}\left(\Psi\right)$. Concerning Eq.(\ref{foe1}) take its complex conjugate and rotate it, $U\left(\vec \sigma \cdot \vec D \Psi \right)^*=0$. It follows from this global transformation that  $U\vec \sigma^* U^{\dagger}\cdot U\vec D^* \Psi^*=\vec \sigma \cdot \vec D(-\vec A) \Psi'=0$. We reach the conclusion that the topological equations naturally break the time-reversal symmetry since they present two independent sets of solutions, namely, ($\Psi,\vec h)$ and $(\Psi',-\vec h)$.

Under the preset scenario it is easy to conclude for the existence of topological solutions according to the following argument. Consider the case of no applied magnetic field and yet the presence of  circulating supercurrents, volumetric between the layers and superficial within the layers, that establish a spatial local magnetic field $\vec h$.
Assume the presence of closed magnetic field stream lines that pierce twice a given layer such that the magnetic field component parallel to the layer flips direction from one side to the other of this layer. Looking from just one side of this layer one sees that a given stream line has a fountain and a sinkhole in this layer. The spatial arrangement of such closed loops is such that the sinkhole of all magnetic field stream lines are concentrated into a few points, the skyrmion cores, whereas the fountains are not necessarily concentrated and in fact are scattered within the unit cell. Recall that this intricate magnetic field arrangement due to the volumetric and superficial supercurrents also results in constant charge rate passing through the layers and creating positive and negative spots within a layer. This results in a highly inhomogeneous state with free energy higher than that of the homogeneous state and so expected to decay. However this does not happen, the state remains stable thanks to its topological properties. Integration over a single layer, chosen at $x_3=0$,
\begin{eqnarray}\label{skyrmion}
Q= \frac{1}{4\pi}\int_{x_3=0^+} \big (\frac{\partial \hat
h}{\partial x_1} \times \frac{\partial \hat h}{\partial x_2}
\big)\cdot \hat h \; d^2x,
\end{eqnarray}
where $\hat h = \vec h/\vert \vec h \vert$, reveals that this inhomogeneous solution has $Q\neq 0$, whereas the homogeneous solution has $Q=0$. This is the skyrmion state and each $Q$ state belongs to a different topological class. Clearly the time reversal symmetry, ($\vec h \rightarrow -\vec h$), is broken by the skyrmions. We find that the topological number $Q$ counts the number of skyrmion cores within the unit cell. Thus the skyrmions are magnetic excitations~\cite{marcelo11} with a core that establishes a well defined sense of rotation in the cell, and for this reason they are also chiral solutions~\cite{takashi12}. The superficial current $\boldsymbol{J_s}$ is very strong within the core as compared to the rest of the cell, where it is weak. At the center of the skyrmion core the rotation ceases. The unique sense of flow set by the core makes the skyrmion state break the time-reversal
symmetry. This preferred chirality of the skyrmions should rotate
circularly polarized light passing through the layers and lead to
the dichroism observed below the pseudogap line\cite{kaminski02,xia08}.

\textbf{The layered solution}. --
Consider the solution of Eqs.(\ref{foe1}) and (\ref{foe2}) for a stack of layers under the simplifying assumption that all layers are identical in the very special limit of a weak $\vec h$ field since this is the interesting physical limit to be treated.
As it is well known NMR/NQR~\cite{strassle08, strassle11}
and $\mu$SR~\cite{macdougall08,sonier09} experiments set a very
restrictive limit to {\it the maximum magnetic field} inside the cuprates, which cannot be larger than $\sim 7 \, \mbox{to} \, 0.7\,\mbox{G}$.
In such a case the solution can be found
recursively, namely, firstly $\Psi$ is obtained from Eq.(\ref{foe1})
in the absence of $\vec h$, and, next, $\vec h$ is determined from
Eq.(\ref{foe2}) using the previously obtained $\Psi$. The smallness
of $\vec h$ dismisses the requirement of further iterations of the
topological equations, such that it becomes enough to solve Eq.(\ref{foe1}) as $\vec{\sigma}\cdot\vec{\nabla}\Psi=0$.
This solution has been obtained elsewhere~\cite{cariglia14} and for a single layer at $x_3=0$ is given by,
\begin{eqnarray}\label{sls}
\Psi = \sum_{\vec k \neq 0} c_{\vec k} \; e^{-k\vert x_3 \vert}
e^{i\vec k \cdot \vec x} \left(
\begin{array}{c} 1 \\ -i\frac{k_{+}}{k}\frac{x_3}{\vert x_3 \vert}, \end{array}\right)
\end{eqnarray}
where $k_\pm = k_1 \pm i k_2$ and $k \equiv \vert\vec k\vert$, $\vec k = k_1\hat x_1+k_2\hat x_2$. Notice that the up and down components satisfy $\psi_u(0^{-})=\psi_u(0^{+})$ and $\psi_d(0^{-})=-\psi_d(0^{+})$, meaning that they correspond to symmetric and antisymmetric modes across the layer of zero thickness. This discontinuous behavior across the layer is necessary because what takes place immediately above and below the layer is very different. From this solution the multi-layer solution valid for $0<x_3<d$ is straightforwardly obtained,
\begin{eqnarray}\label{mls}
\Psi = \sum_{\vec k \neq 0} c_{\vec k} \; \frac{e^{i\vec k \cdot
\vec x}}{\sinh\left(k d/2\right)} \left(
\begin{array}{c} \cosh \left[k\left(x_3-d/2\right) \right] \\ i\frac{k_{+}}{k} \sinh \left[k\left(x_3-d/2\right)
\right]\end{array}\right).\nonumber \\ \label{op3d}
\end{eqnarray}
In both cases the superficial current $\vec
J_s$ in the layers is immediately determined from,
\begin{eqnarray}\label{curr-sup}
 \frac{\vec J_s}{2c\mu_B} =  \Psi^\dag (0^+)
\sigma_2\Psi(0^+) \hat x_1 -\Psi^\dag (0^+)
\sigma_1\Psi(0^+) \hat x_2,
\end{eqnarray}
where,
\begin{eqnarray}\label{sig1}
&&\Psi^\dag (0^+) \sigma_1\Psi(0^+) =-\sum_{\vec {k'},\, \vec k \neq 0}c^{*}_{\vec {k'}}c_{\vec k}\,e^{i \left(\vec k-\vec {k'}\right) \cdot \vec x}\cdot\nonumber\\
&&\cdot \left[i\frac{k_+}{k}\frac{1}{\tanh\left(\frac{k'd}{2}\right)} -i \frac{k'_{-}}{k'}\frac{1}{\tanh\left(\frac{k d}{2}\right)}\right],
\end{eqnarray}
and,
\begin{eqnarray}\label{sig2}
&&\Psi^\dag (0^+) \sigma_2\Psi(0^+) =\sum_{\vec {k'},\, \vec k \neq 0}c^{*}_{\vec {k'}}c_{\vec k}\,e^{i \left(\vec k-\vec {k'}\right) \cdot \vec x}\cdot\nonumber\\
&&\cdot \left[\frac{k_+}{k}\frac{1}{\tanh\left(\frac{k'd}{2}\right)} + \frac{k'_{-}}{k'}\frac{1}{\tanh\left(\frac{k d}{2}\right)}\right].
\end{eqnarray}
The order parameter $\Psi$ of Eq.(\ref{mls}) is intrinsically inhomogeneous since $\vec k \neq 0$. We choose to study here a periodic structure
characterized by a unit cell with sides $L_1$ and $L_2$. Therefore $k_i=2\pi
n_i/L_i$, $i=1,2$, where $n_1$ and $n_2$ are integers. Thus the
volumetric cell has volume $V=Ad$, $A=L_1L_2$ being the rectangular
area within the layer where we find $Q$ skyrmions.
Albeit its complexity, the superficial supercurrent in the unit cell, which has area $A$ and a perimeter $P$, satisfies some simple properties:
\begin{enumerate}
\item[i)] Null average supercurrent within the unit cell: $\int_{A} \boldsymbol{J_s}
d^2 x =0$;
\item[ii)] No net supercurrent circulation at the edge of the unit cell: $\oint_{P} \boldsymbol{J_s}\cdot d\vec l=0$; and
\item[iii)] No in and out of the unit cell supercurrent flow: $\oint_{P}
\boldsymbol{J_s}\cdot \hat n \, d l=0$ where $\hat n \cdot d\vec
l=0$.
\end{enumerate}
It is straightforward to check that $\int_{A} \boldsymbol{J_s}
d^2 x =-4c\mu_B A \sum_{\vec k} \vec k \vert c_{\vec
k}\vert^2/\tanh(k d/2)$ using that $\int_{A} \exp{[i(\vec k-\vec
{k'})\cdot \vec x]} d^2x=\delta_{\vec k,\vec {k'}}A$. This summation
vanishes provide that the coefficients satisfy
$|c_{-k_1,k_2}|^2=|c_{k_1,k_2}|^2$, and $|c_{k_1,-k_2}|=|c_{k_1,k_2}|^2$, and so
we find that (i) is valid under this condition. We use Stoke's
theorem, $\oint_{P} \boldsymbol{J_s}\cdot d\vec l=\int_{A} \hat x_3
\cdot(\vec \nabla \times \boldsymbol{J_s})d^2 x$, and Gauss'
theorem, $\oint_{P} \boldsymbol{J_s}\cdot \hat n \, d l=\int_{A}
\vec \nabla \cdot \boldsymbol{J_s} d^2 x$ to find that assertions
(ii) and (iii) are true, respectively, because $\int_{A} \nabla_i
{J_s}_j d^2 x=0$ for any $j=1,2$ since $\left(k_i-{k'}_i\right)\int_{A}
\exp{[i(\vec k-\vec {k'})\cdot \vec x]} d^2x=0$ for $i=1,2$. An
important and direct consequence of (ii) is that there is no net
charge rate entering or exit the cell,
\begin{eqnarray}\label{avsig}
\int_{A} \frac{\partial\sigma}{\partial t}d^2x=0,
\end{eqnarray}
according to Eq.(\ref{sigma}).


\textbf{The Ginzburg-Landau theory}. -- We show here that the topological equations do solve the Ginzburg-Landau variational equations for the special choice of temperature $T=T_c=T^*$ under the approximation of a very weak local magnetic field. There is no applied magnetic field but there is a circulating supercurrent that creates this very small magnetic field. This solution corresponds to a lattice of skyrmions that we claim to be solution for the GL theory in this particular temperature.
Recall that the GL theory is an order parameter expansion valid for temperatures near to the critical one where the order parameter is supposed to be small since the superconducting state is at the brink of disappearance. Obviously at the transition temperature itself the order parameter should be very small indeed. Landau's argument is that in the $T_c$ neighborhood powers of the order parameter higher than four can be safely neglected.
Firstly, consider the case of the traditional Ginzburg-Landau theory, without the presence of an external field, whose Gibbs free energy is the sum of three terms, namely, the kinetic, the condensate and the field density energies,
\begin{eqnarray}
&&F=F_k+F_c+F_f, \\
&&F_k = \langle \frac{\vert \vec{D}\psi \vert^2}{2m} \rangle, \label{fk0}\\
&&F_c = \langle-\alpha_0|\psi|^2 +\frac{1}{2}\beta |\psi|^4 \rangle,\; \mbox{and}, \label{fc0}\\
&&F_f = \langle\frac{\vec h^2}{8\pi}\rangle,\label{ffield0}
\end{eqnarray}
where $\langle \cdots \rangle \equiv \int \left
(\cdots \right)d^3x/V$ and $V$ is the bulk volume, $m$ is the Cooper pair mass, $\alpha_0=c_0(T_c-T)$, $c_0>0$, and $\beta>0$.
For $T=T_c$ the condensate becomes positive, $F_c\ge 0$, and so all of the contributions to the free energy are positive, since $F_k\ge 0$ and  $F_f\ge 0$ hold for any temperature. Thus the lowest energy state has $F=0$, and  corresponds to the homogeneous state $\psi=0$  without any local magnetic field present, $\vec h = 0$.
However the situation becomes far more complex in case of the two-component order parameter GL theory given by,
\begin{eqnarray}
&&F=F_k+F_c+F_f, \\
&&F_k = \langle \frac{\vert \vec{D}\Psi \vert^2}{2m} \rangle, \; \mbox{and}, \label{fkin0}\\
&&F_c = \langle-\alpha_0|\Psi|^2-\vec \alpha \cdot \Psi^{\dag}\vec
\sigma\Psi +\frac{1}{2}\Psi^* \cdot \Psi^* \cdot \beta \cdot\Psi \cdot \Psi \rangle.\nonumber \\ \label{fcond0}
\end{eqnarray}
The condensate energy density is assumed to be the most general one with no extra assumptions other than its own stability. The second order term is the most general one and contains four independent  parameters, $\alpha_0$, $\alpha_3$ and two other ones, $\vec \alpha_{\parallel}$, where parallel means to the layers as we shall see here. The fourth order term must be real and  positive to warrant stability of the condensate energy: $\Psi^* \cdot \Psi^* \cdot \beta \cdot\Psi \cdot \Psi\equiv \beta_{abcd}\psi^{*}_a\psi^{*}_b \psi_c\psi_d> 0$, where the indices
$a,b,c,d$ run over $u$ and $d$ in the most general  tensor
$\beta_{abcd}$, which contains the required symmetry to render the fourth order term also positive. However the presence of two critical temperatures introduces new features into the theory.
The presence of two critical temperatures restricts, according to the above argument, the validity of this GL free energy expansion to the temperature range $T \approx T^*$, and $T \approx T_c$ where the order parameter is expected to be small. Consequently the validity of this GL free energy expansion is also limited to situations such that $T^* \approx T_c$. Assume that the pseudogap and the superconducting transition temperatures can be associated to $T^*$ and $T_c$, respectively. Thus from the point of view of the temperature versus doping diagram, the present arguments restrict a GL free energy expansion to the top of the superconducting dome where these two lines cross each other.
The temperature $T=T_c=T^*$ corresponds to $\alpha_0+\alpha_3=0$
and $\alpha_0-\alpha_3=0$,  since
$\alpha_0|\Psi|^2+\vec \alpha \cdot \Psi^{\dag}\vec
\sigma\Psi= \left(\alpha_0+\alpha_3 \right) |\psi_u|^2+ \left(\alpha_0-\alpha_3 \right) |\psi_d|^2+\vec \alpha_{\parallel} \cdot \Psi^{\dag}\vec\sigma_{\parallel}\Psi$. We also assume that at this crossing
temperature that $\vec \alpha_{\parallel}=0$, where parallel is
associated to the direction along the layers by choice of coordinate system. We shall see that the topological solution automatically satisfies that $\langle \Psi^{\dag}\vec \sigma_{\parallel}\Psi \rangle=0$.
Similarly to the one-component GL theory  the free energy of the two-component case also becomes a sum of three positive terms in this
special temperature. Thus one naturally expects that its fundamental
state has $F=0$ which corresponds to $\Psi=0$ and $\vec h = 0$.
Indeed this is the case, but we shall show here that there is an
excited inhomogeneous state above this homogeneous state such that $\Psi \ne 0$ and $\vec h \ne 0$. This is the skyrmion state, made stable because of its topological properties. The previous solution of the two-component GL theory, given by Eq.(\ref{op3d}), obtained under the only assumption that the order parameter is small, fact that defines an expansion parameter $\varepsilon$, namely, $\Psi=O(\varepsilon)$.
However the order parameter is not dimensionless  since $\Psi^{\dag}\Psi$ is a density and has the dimension of $1/V$, where $V$ is the volume.
Therefore we seek to determine $\varepsilon \propto 1/\sqrt{V}$ and leave to show elsewhere a more careful analysis that treats the present expansion in terms of a dimensionless order parameter.
We show that the topological equations, provide a solution of the variational equations to order $O(\varepsilon^3)$ for the temperature $T=T_c=T^*$.

The variational second order equations of the two-component order parameter theory are given by,
\begin{eqnarray}
&& \frac{\vec D^2 \Psi}{2m}=\alpha_0\Psi+\vec \alpha \cdot \sigma\Psi -  \left (\Psi^*\cdot \beta \cdot \Psi \right )\cdot\Psi \label{gleq0}\\
&& \vec \nabla \times \vec h = \frac{4\pi}{c} \vec J, \, \vec J =
\frac{q}{2m}\left (\Psi^{\dag}\vec D \Psi + c.c. \right)\label{ampere0}
\end{eqnarray}
The cubic order term in the order parameter means, for instance, that the "d" component of $\left (\Psi^* \cdot \beta \cdot\Psi \right ) \cdot \Psi$ is  $\beta_{abcd}\psi^{*}_a\psi^{*}_b \psi_c$.

The keystone of the present approach is the existence of a dual formulation of the kinetic energy density ~\cite{alfredo13,cariglia14} given by,
\begin{eqnarray}\label{kin2}
&&F_k=\langle \frac{1}{2m}\left\vert\vec{\sigma}\cdot\vec{D}\Psi\right\vert^2+\mu_B \vec{h} \cdot\Psi^\dag\vec{\sigma}\Psi- \nonumber
\\&&-\frac{\hbar}{4m}\vec \nabla\left[\Psi^\dag\left(\vec \sigma \times
\vec D\right)\Psi+c.c. \right]\rangle.
\end{eqnarray}
While the original formulation of the kinetic energy leads to the above standard formulation of the variational equation, the dual one leads to an equivalent, but distinct formulation, given by,
\begin{eqnarray}
&& \frac{1}{2m}\left( \vec{\sigma}\cdot\vec{D}\right)^2 \Psi= \nonumber \\
&& -\mu_B \vec h \cdot \vec \sigma \Psi+ \alpha_0\Psi+\vec \alpha \cdot \vec \sigma\Psi -  \left (\Psi^*\cdot \beta \cdot \Psi \right )\cdot\Psi \label{gleq1} \\
&& \vec \nabla \times \left(\vec h + 4\pi\mu_B\Psi^\dag  \vec{\sigma}\Psi \right) = \nonumber \\
&& =\frac{2\pi q }{mc}\left[\Psi^\dag\vec \sigma
\left (\vec{\sigma}\cdot\vec{D}\Psi\right )+ c.c.\right ]. \label{ampere1}
\end{eqnarray}
For instance, to show that the GL equation admits this
twofold formulation, namely, that Eqs.(\ref{gleq1}) and
(\ref{gleq0}) are equivalent, just use that $\vec
D^2\Psi=I\delta_{ij}D_iD_j\Psi$, and that
$I\delta_{ij}=\sigma_i\sigma_j-i\epsilon_{ijk}\sigma_k$, where $I$
is the two by two identity matrix, $\epsilon_{ijk}$ is the totally
anti-symmetric Levi-Civita tensor, and so, the local magnetic field
is $h_k=-i\epsilon_{ijk} D_i D_j$.

Next we show that the second order variational equations are solved by the FOEs until order lower than $O(\varepsilon^3)$. The right side of the GL equation, given by Eq.(\ref{gleq1}), is of order
$O(\varepsilon^3)$, since $\Psi=O(\varepsilon)$ and $\vec h=O(\varepsilon^2)$. The terms of order $O(\varepsilon)$ in the right side are considered to vanish at $T=T_c=T^*$, and the ones of order $O(\varepsilon^3)$ are negligibly small and can be approximated to zero, while the left side vanishes by virtue of the topological equation given by Eq.(\ref{foe1}). Remarkably Amp\`ere's law, given by Eq.(\ref{ampere1}), is exactly solved by the topological equations.

The free energy is of order $O(\varepsilon^2)$  for $T=T_c=T^*$. To show this firstly write Eq.(\ref{foe1}) as,
\begin{eqnarray}
\vec{D}\, \Psi= i\vec{\sigma}\times \vec{D}\, \Psi. \label{foe1b}\\
\end{eqnarray}
Then the kinetic energy density of Eq.(\ref{kin2}) becomes,
\begin{eqnarray}\label{kin2b}
F_k=\langle \frac{\hbar^2}{4m}\nabla^2 \vert\Psi\vert^2 - \frac{1}{4\pi} \vec h^2 \rangle.
\end{eqnarray}
by use of Eq.(\ref{foe2}). The condensate energy density, Eq.(\ref{fcond0}), for $T=T_c=T^*$ features $F_c=O(\varepsilon^4)$ and  the
field energy, Eq.(\ref{ffield0}), is also $F_f=O(\varepsilon^4)$.
Therefore the total energy has only one term of order
$O(\varepsilon^2)$, and becomes,
\begin{eqnarray}\label{ftot}
F=\langle \frac{\hbar^2}{4m}\nabla^2 \vert\Psi\vert^2 \rangle+ O(\varepsilon^4).
\end{eqnarray}
Interestingly the above $O(\varepsilon^2)$  term is a surface one that does not vanish thanks to the distinct behavior of the skyrmion solution infinitesimally above and below a layer. In fact this term is responsible for the gap of the inhomogeneous state above the homogeneous one because it reaches a constant value in its infrared limit (small k)~\cite{cariglia14}.
The free energy  follows from Eqs.(\ref{op3d})
and (\ref{ftot}), which give that,
\begin{eqnarray}\label{gap}
F=\frac{(h/d)^2}{\pi^2m}\sum_{\vec k \neq 0} \vert c_{\vec
k}\vert^2 \frac{kd/2}{\tanh\left(kd/2\right)}+ O(\varepsilon^4).
\end{eqnarray}
Thus we have proven that at least for $T=T_c=T^*$ the topological equations solve the variational ones which means that the free energy can be determined from the above expression.

Nevertheless our criterion for abandoning terms of order $O(\varepsilon^3)$ and higher in the variational equations has introduced a handicap into the problem. The normalization parameter $\varepsilon$, so far just
assumed to be small, cannot be determined by keeping just terms below order $O(\varepsilon^3)$. The local magnetic field, according to
Eq.(\ref{foe2}), and the free energy, Eq.(\ref{ftot}) are both of
order $O(\varepsilon^2)$, but there is no scheme to determine this
parameter. Indeed the $\varepsilon$ parameter should be determined by the condensate energy, which here was completely abandoned because of it smallness. Therefore we introduce
an external phenomenological criterion  to define $\varepsilon$, which corresponds to the
knowledge of the local magnetic field inside the superconductor. In
some sense such knowledge is a way to phenomenologically include the
residual higher order terms $O(\varepsilon^4)$ present in the free
energy that were abandoned. Therefore we shall use this to argue through Eq.(\ref{foe2}) that $\varepsilon \sim \sqrt{h_{exp}}$.


\textbf{States of angular momentum}. -- The topological equations have a degenerate set of solutions, which means that these equations do leave room in parameter space for further minimization of the free energy minimization.
They present an undetermined number of solutions as seen in the coefficients $c_{\vec k}$ of the order parameter which are not determined by the topological equations.
This freedom also shows that for $T=T_c=T^*$ there is a large degeneracy in the problem in case terms of order $O(\epsilon^3)$ in the variational equations are neglected.
To explicitly calculate the $\partial\sigma / \partial t$ of the skyrmion state we need to know the coefficients $c_{\vec k}$ in Eq.(\ref{op3d}) and do it here so to represent an order parameter with a definite angular momentum perpendicular to the layers.
\begin{eqnarray}\label{j3e}
J_3\Psi_m=\hbar\left(m+\frac{1}{2}\right)\Psi_m, \, \mbox{where} \, J_3 = l_3+\frac{\hbar}{2}\sigma_z,
\end{eqnarray}
$l_3=x_1p_2-x_2p_1$, whose position representation is $p_i=(\hbar/i)(\partial/\partial x_i)$, and the wave number representation is $x_i=i(\partial/\partial k_i)$ and $p_i=\hbar k_i$.
Therefore it holds that $l_3 c_{\vec k} = \hbar m c_{\vec k}$ whose solution is $c_{\vec k}= \left(k_{+}/k\right)^m$.
For simplicity we take the order parameter of Eq.(\ref{op3d}) with only the lowest fourier terms  included, namely, $n_i
= −1, 0, 1$, $i = 1,2$ ($n_1 = n_2 = 0$ is excluded). Then one obtains that,
\begin{widetext}
\begin{eqnarray}
&&\Psi_{m}=\varepsilon{\frac{2}{\sinh(\frac{gd}{2})}\left( \begin{array}[c]{cc}
\left[e^{i\frac{m\pi}{2}}\cos\left(g x_1-\frac{m\pi}{2}\right)+e^{i m\pi}\cos\left(g x_2-\frac{m\pi}{2}\right)\right]\cosh\left( g\overline{x}_3 \right)\\
-i\left[e^{i\frac{\overline{m}\pi}{2}}\cos\left(g x_1-\frac{\overline{m}\pi}{2}\right)+e^{i \overline{m} \pi}\cos\left(g x_2-\frac{\overline{m}\pi}{2}\right)\right]\sinh\left( g\overline{x}_3 \right)
\end{array} \right)}\nonumber\\
&&+\varepsilon\frac{2}{\sinh(\frac{\sqrt{2}gd}{2})}\left( \begin{array}[c]{cc}
\{e^{i\frac{3m\pi}{4}}\cos\left[g(x_2+x_1)-\frac{m\pi}{2}\right]+e^{i\frac{5m\pi}{4}}\cos\left[g(x_2-x_1)-\frac{m\pi}{2}\right]\}\cosh\left(\sqrt{2}g\overline{x}_3 \right)\\
-i\{e^{i\frac{3\overline{m}\pi}{4}}\cos\left[g(x_2+x_1)-\frac{\overline{m}\pi}{2}\right]+e^{i\frac{5\overline{m}\pi}{4}}\cos\left[g(x_2-x_1)-\frac{\overline{m}\pi}{2}\right]\}\cosh\left(\sqrt{2}g\overline{x}_3 \right)
\end{array} \right),\nonumber \\
\label{psipm}
\end{eqnarray}
\end{widetext}
where $\overline{m}\equiv m+1$, $g\equiv 2\pi/L$ and  $\overline{x}_3 \equiv x_3-d/2$. Then from Eq.(\ref{currents}) and $\partial \sigma/\partial t = - \vec \nabla \cdot \vec J_s$, one obtains that,
\begin{eqnarray}\label{sigma1}
&& \frac{\partial \sigma_{m}}{\partial t}= -16c\mu_{B}g\varepsilon^2 \alpha_{m} \Sigma(x_1,x_2), \, \mbox{where} \\
&& \Sigma(x_1,x_2) \equiv \sin (g x_1) \sin (g x_2) \left[\cos (g x_1)-\cos (g x_2)\right ]. \nonumber
\end{eqnarray}
Notice that only the multiplicative coefficient carries information about the angular momentum state, that is,
\begin{eqnarray}
\alpha_{m}\equiv p_{m} \coth \left(\frac{d g}{2}\right)-q_{m} \coth \left(\frac{d g}{2 \sqrt{2}} \right),
\end{eqnarray}
where the coefficients $p_m$, $q_m$ are defined in table \ref{table1}.
\begin{table}[b]
\caption{The
coefficients $\alpha_{m}$ for $m=-4,\ldots,+4$ are listed in this table.} \centering
\begin{tabular}{|c|c|c|}
\hline m & $p_m$ & $q_m$\\
\hline -4 & $\sqrt{2}$ & $2$  \\
\hline -3 & $-4$ & $-3\sqrt{2}$  \\
\hline -2 & $3\sqrt{2}$  & $4$  \\
\hline -1 & $-2$ & $-\sqrt{2}$  \\
\hline 0  & $-\sqrt{2}$ & $-2$  \\
\hline 1  & $4$ & $3\sqrt{2}$  \\
\hline 2  & $-3\sqrt{2}$ & $-4$  \\
\hline 3  & $2$ & $\sqrt{2}$  \\
\hline 4  & $\sqrt{2}$ & $2$  \\
\hline
\end{tabular}
\label{table1}
\end{table}
The value of $\varepsilon$ is determined from the assumption that the calculated mean value of the local magnetic field corresponds to the experimental threshold, namely, $h_{exp}=\vert\langle \vec h \rangle \vert$. Therefore assuming the angular momentum states previously discussed we obtain that~\cite{alfredo14},
\begin{eqnarray}\label{hmax0}
h_{exp}=16\pi\mu_{B}\varepsilon^2\left[\frac{1}{\sinh^2\left(\frac{\pi d}{L}\right)}+\frac{1}{\sinh^2\left(\frac{\sqrt{2}\pi d}{L}\right)}\right]
\end{eqnarray}
The present predictions for the inhomogeneous state gap are based on the  selection of parameters, namely, the experimental threshold for the local field, $h_{exp}=0.01\; \mbox{Gauss}$, and the ratio $d/L=0.75$. For the latter we have in mind the compound $YBa_2Cu_3O_{7-0.08}$ as this material presents the checkerboard pattern~\cite{zahirul04} with $L=4a=1.6 \; nm$, where the crystallographic cell has size $a = 0.4 \; \mbox{nm}$ and $d = 1.2 \; \mbox{nm}$. Notice that in the present model there is no commensurability with the underlying crystallographic structure such that the ratio $L/a$ can be any. The only relevant concern of the present model is whether the existence of skyrmions sets some limits on the ratio $L/d$. Indeed we have found an upper bound for $L/d$ much above the above taken value~\cite{alfredo14} of  $d/L=0.75$. In convenient units the Bohr magneton is $\mu_B=9.2 \; \mbox{Gauss}\cdot\mbox{nm}^3$, and then we obtain that~\cite{alfredo14},
\begin{equation}
\varepsilon^2=5.3\times 10^{-4} \; \mbox{nm}^{-3}.
\end{equation}
As previously discussed~\cite{cariglia14} the inhomogeneous state gap of Eq.(\ref{gap}) becomes $F = 0.5 \; \mbox{meV.nm}^{-3}$. A dimensional analysis shows that the charge density rate is controlled by the mean magnetic field and the size of the tetragonal lattice, $L$:
\begin{equation}
\frac{\partial \sigma}{\partial t}\sim \frac{h_{exp}c}{L}
\end{equation}
This follows from the following argument. According to  Eq.(\ref{currents}) $J_s \sim c \mu_{B}\varepsilon^2$ since $\Psi \sim \varepsilon$. Thus Eq.(\ref{sigma}) sets that $\partial\sigma / \partial t \sim c \mu_{B}\varepsilon^2/L$ since $\nabla \sim 1/L$, as it becomes evident in Eq.(\ref{sigma1}). From the other side Eq.(\ref{hmax0}) sets that $h_{exp}\sim \mu_{B}\varepsilon^2$ which leads to the above result. Under these values one obtains that,
\begin{eqnarray}
\frac{\partial \sigma}{\partial t} \sim 10^{-7} \, \mbox{A/nm$^2$}.
\end{eqnarray}
This is the estimated charge density rate which achieves positive and negative values within the cell such that its average value vanishes as shown in Eq.(\ref{avsig}).

From this value we also estimate the time rate
that pairs enter the layers, based on an extra assumption beyond the scope of the present model. We assume that  pairs cross a
layer, $q \sim 3.2\,10^{-19} \, \mbox{C}$, within an area of $1.0 \, \mbox{nm}^2$ with some fixed frequency $f$. Notice that they do it in both senses although in different spots.
This periodic entrance and exit defines a rate, and so a natural
frequency $f$ to the layered system,  obtained by equating $ q
f/\mbox{nm}^2 = 10^{-7} \, \mbox{A/nm$^2$}$, which gives
\begin{equation}\label{freq}
f \sim 0.3 \, 10^{12}\, \mbox{Hz}.
\end{equation}
Interestingly this number falls in the same order of magnitude of the
Josephson effect between layers in the cuprates~\cite{kadowaki13} which are in the THz regime.

\textbf{The charge density wave}. -- In the present magnetostatic model there is charge crossing a layer at constant rate that we interpret as the origin of a charge density wave. Recently G. Ghiringhelli et al.~\cite{ghiringhelli12}  have
proposed the presence of a charge density wave in the CuO$_2$ layers of the cuprates  using resonant soft x-ray scattering. This
two-dimensional charge density wave in the underdoped compound
YBa$_2$Cu$_3$O$_{6+x}$ with an incommensurate periodicity that sets a
tetragonal lattice because it is found to exist in orthogonal
directions, namely, along and perpendicular to the so-called CuO
chains. Interestingly they find that this structure holds both
above and below $T_c$. The present model of two-dimensional  layers embedded in a kind of metallic medium displays an inhomogeneous charge distribution in the layers. However from a three-dimensional perspective of the material there is no charge accumulation in any point since both the volumetric and the superficial supercurrent render the total supercurrent divergenceless.
Although the charges are not static they are crossing the layers at constant rate and in so doing they define positive and negative spots according to their entrance and exit in the layers. This follows from $\partial \sigma /\partial t$ which was previously calculated in the layers.
We address here the question of the multipole moment of this charge density in the layer. To achieve this goal we briefly review a few aspects of multipole expansion suitable for our analysis. The electrostatic energy density $U$ associated to a charge density $\sigma(\vec x)$ in presence of an applied electrostatic potential $V(\vec x)$ is given by,
\begin{eqnarray}
U = \int_{A} \frac{d^2x}{A} \sigma(\vec x) V(\vec x)
\end{eqnarray}
Obviously here we are considering that we are looking at a fixed time window such that $\Delta t << 1/f$, f given Eq.(\ref{freq}):
\begin{eqnarray}
\sigma \approx \frac{\partial \sigma}{\partial t}\Delta t.
\end{eqnarray}
Expanding the electrostatic potential around a point $\vec x = 0$ within the area $A$ gives that,
\begin{eqnarray}
&& U = QV(0)+\sum_{N=1}^{\infty}\frac{1}{N!}Q_{i_{1}\,i_{2}\cdots i_{N}}V_{i_{1}\,i_{2}\cdots i_{N}}(0) \\
&& Q_{i_{1}\,i_{2}\cdots i_{N}}\equiv \int_{A} d^2x \sigma(\vec x)\, x_{i_{1}}x_{i_{2}}\cdots x_{i_{N}} \\
&& V_{i_{1}\,i_{2}\cdots i_{N}}(0) \equiv \frac{\partial^N V(\vec x)}{\partial x_{i_{1}}\partial x_{i_{2}}\cdots\partial x_{i_{N}}}|_{\vec x=0}
\end{eqnarray}
where $Q=\int_{A}d^2x \sigma(\vec x)$ is the total charge and $V(0)$ the potential at this selected origin. The tensors $Q_{i_{1}\,i_{2}\cdots i_{N}}$ are the multipole moments of this charge distribution. A way to calculate these tensors is simply to obtain the Fourier transform of the charge density and then expand it in powers of the wave number:
\begin{eqnarray}
\int_{A} d^2x e^{i\vec k \cdot \vec x}\sigma(\vec x) = Q+\sum_{N=1}^{\infty}\frac{i^N}{N!}Q_{i_{1}\,i_{2}\cdots i_{N}}k_{i_{1}}k_{i_{2}}\cdots k_{i_{N}}
\end{eqnarray}
Next we apply the above ideas to the definite angular momentum states and obtain the Fourier transform of  the charge density rate given by Eq.(\ref{sigma1})
\begin{widetext}
\begin{equation}
\int_{A} e^{i\vec k \cdot \vec x} \sigma_m \, d^2x=-c\mu_{B}\alpha_m\frac{48 g^5 \left(-1+e^{\frac{2 i \pi k_1}{g}}\right) \left(-1+e^{\frac{2 i \pi k_2}{g}}\right) \left(k_1^2-k_2^2\right)}{\left(4 g^4-5 g^2 k_1^2+k_1^4\right) \left(4 g^4-5 g^2 k_2^2+k_2^4\right)}\Delta t
\end{equation}
\end{widetext}
Expanding in powers of wave number gives that,
\begin{equation}
\int_{A} e^{i\vec k \cdot \vec x} \sigma_m d^2x=-c\mu_{B}12\pi^2\alpha_m\frac{1}{g^7}(k_2k_1^3-k_1k_2^3)\Delta t+O(k^5),
\end{equation}
fact that configures an hexadecapole charge distribution since  the first non-vanishing moments are,
\begin{equation}
Q_{1112}=Q_{2221}=-c\mu_{B}12\pi^2\alpha_m\frac{1}{g^7}\Delta t.
\end{equation}

Fig.~\ref{mstates} displays the superficial current density within the unit cell for $m=0,1,2,\mbox{and}\;3$ associated to $J_3$ given by Eq.(\ref{j3e}). The skyrmion cores can be seen in these plots and correspond to $Q=+1,+2,-2,-1$, respectively. Interestingly the $m=0$ state has a core sitting at the corner and the figure display one fourth of it sitting at each corner. This also holds for the $m=1$ state, whose core is at the corner but has double charge. The remaining two states, $m=2,3$ can be interpreted from the two previous cases, $m=1,0$ such that the circulation in the middle is strengthened in the center and weakened at the corner, which reverts the skyrmion number.

Fig.~\ref{charge} shows the charge density rate crossing the unit cell and generating a hexadecapole moment. Positive and negative charged spots  represent the entrance and exit of the volumetric supercurrent in the unit cell at constant rate. Interestingly all the $m$ states have the same charge density rate spatial distribution that only differs by the amplitude, according to the coefficients of Table \ref{table1}. Notice that the $m=2$ state has opposite signal in comparison with the other $m>0$ states. Interestingly the position of the skyrmion cores, shown in Fig.~\ref{mstates},  and the positive and negative spots of the charge density rate, shown in Fig.~\ref{charge}, are uncorrelated.

\textbf{Conclusion}. --
We have shown that the two-component order parameter theory describes layers constantly crossed by the supercurrent forming positive and negative charged spots in the condensate. Therefore this condensate is an inhomogeneous state with a gap above the ground state that produces a local magnetic field below the threshold of experimental observation and without the presence of an applied external field. This state is topologically stable and given by a lattice of skyrmions that display an elaborate pattern of volumetric and superficial currents circulating through the stack of layers that breaks time reversal symmetry.

\textbf{Acknowledgments}:
Mauro M. Doria and Alfredo. A. Vargas-Paredes acknowledge the Brazilian agency CNPq for financial support.

\bibliography{reference-sust}

\begin{thebibliography}{29}%
\makeatletter
\providecommand \@ifxundefined [1]{%
 \@ifx{#1\undefined}
}%
\providecommand \@ifnum [1]{%
 \ifnum #1\expandafter \@firstoftwo
 \else \expandafter \@secondoftwo
 \fi
}%
\providecommand \@ifx [1]{%
 \ifx #1\expandafter \@firstoftwo
 \else \expandafter \@secondoftwo
 \fi
}%
\providecommand \natexlab [1]{#1}%
\providecommand \enquote  [1]{``#1''}%
\providecommand \bibnamefont  [1]{#1}%
\providecommand \bibfnamefont [1]{#1}%
\providecommand \citenamefont [1]{#1}%
\providecommand \href@noop [0]{\@secondoftwo}%
\providecommand \href [0]{\begingroup \@sanitize@url \@href}%
\providecommand \@href[1]{\@@startlink{#1}\@@href}%
\providecommand \@@href[1]{\endgroup#1\@@endlink}%
\providecommand \@sanitize@url [0]{\catcode `\\12\catcode `\$12\catcode
  `\&12\catcode `\#12\catcode `\^12\catcode `\_12\catcode `\%12\relax}%
\providecommand \@@startlink[1]{}%
\providecommand \@@endlink[0]{}%
\providecommand \url  [0]{\begingroup\@sanitize@url \@url }%
\providecommand \@url [1]{\endgroup\@href {#1}{\urlprefix }}%
\providecommand \urlprefix  [0]{URL }%
\providecommand \Eprint [0]{\href }%
\providecommand \doibase [0]{http://dx.doi.org/}%
\providecommand \selectlanguage [0]{\@gobble}%
\providecommand \bibinfo  [0]{\@secondoftwo}%
\providecommand \bibfield  [0]{\@secondoftwo}%
\providecommand \translation [1]{[#1]}%
\providecommand \BibitemOpen [0]{}%
\providecommand \bibitemStop [0]{}%
\providecommand \bibitemNoStop [0]{.\EOS\space}%
\providecommand \EOS [0]{\spacefactor3000\relax}%
\providecommand \BibitemShut  [1]{\csname bibitem#1\endcsname}%
\let\auto@bib@innerbib\@empty
\bibitem [{\citenamefont {Bednorz}\ and\ \citenamefont
  {M\"uller}(1986)}]{bedmull86}%
  \BibitemOpen
  \bibfield  {author} {\bibinfo {author} {\bibfnamefont {J.}~\bibnamefont
  {Bednorz}}\ and\ \bibinfo {author} {\bibfnamefont {K.}~\bibnamefont
  {M\"uller}},\ }\href {\doibase 10.1007/BF01303701} {\bibfield  {journal}
  {\bibinfo  {journal} {Zeitschrift f\"ur Physik B Condensed Matter}\ }\textbf
  {\bibinfo {volume} {64}},\ \bibinfo {pages} {189} (\bibinfo {year}
  {1986})}\BibitemShut {NoStop}%
\bibitem [{\citenamefont {Sigrist}\ and\ \citenamefont
  {Ueda}(1991)}]{sigrist91}%
  \BibitemOpen
  \bibfield  {author} {\bibinfo {author} {\bibfnamefont {M.}~\bibnamefont
  {Sigrist}}\ and\ \bibinfo {author} {\bibfnamefont {K.}~\bibnamefont {Ueda}},\
  }\href {\doibase 10.1103/RevModPhys.63.239} {\bibfield  {journal} {\bibinfo
  {journal} {Rev. Mod. Phys.}\ }\textbf {\bibinfo {volume} {63}},\ \bibinfo
  {pages} {239} (\bibinfo {year} {1991})}\BibitemShut {NoStop}%
\bibitem [{\citenamefont {Brandt}(1995)}]{brandt95}%
  \BibitemOpen
  \bibfield  {author} {\bibinfo {author} {\bibfnamefont {E.~H.}\ \bibnamefont
  {Brandt}},\ }\href {http://stacks.iop.org/0034-4885/58/i=11/a=003} {\bibfield
   {journal} {\bibinfo  {journal} {Reports on Progress in Physics}\ }\textbf
  {\bibinfo {volume} {58}},\ \bibinfo {pages} {1465} (\bibinfo {year}
  {1995})}\BibitemShut {NoStop}%
\bibitem [{\citenamefont {Bosma}\ \emph {et~al.}(2011)\citenamefont {Bosma},
  \citenamefont {Weyeneth}, \citenamefont {Puzniak}, \citenamefont {Erb},
  \citenamefont {Schilling},\ and\ \citenamefont {Keller}}]{bosma11}%
  \BibitemOpen
  \bibfield  {author} {\bibinfo {author} {\bibfnamefont {S.}~\bibnamefont
  {Bosma}}, \bibinfo {author} {\bibfnamefont {S.}~\bibnamefont {Weyeneth}},
  \bibinfo {author} {\bibfnamefont {R.}~\bibnamefont {Puzniak}}, \bibinfo
  {author} {\bibfnamefont {A.}~\bibnamefont {Erb}}, \bibinfo {author}
  {\bibfnamefont {A.}~\bibnamefont {Schilling}}, \ and\ \bibinfo {author}
  {\bibfnamefont {H.}~\bibnamefont {Keller}},\ }\href {\doibase
  10.1103/PhysRevB.84.024514} {\bibfield  {journal} {\bibinfo  {journal} {Phys.
  Rev. B}\ }\textbf {\bibinfo {volume} {84}},\ \bibinfo {pages} {024514}
  (\bibinfo {year} {2011})}\BibitemShut {NoStop}%
\bibitem [{\citenamefont {Welp}\ \emph {et~al.}(2013)\citenamefont {Welp},
  \citenamefont {Kadowaki},\ and\ \citenamefont {Kleiner}}]{kadowaki13}%
  \BibitemOpen
  \bibfield  {author} {\bibinfo {author} {\bibfnamefont {U.}~\bibnamefont
  {Welp}}, \bibinfo {author} {\bibfnamefont {K.}~\bibnamefont {Kadowaki}}, \
  and\ \bibinfo {author} {\bibfnamefont {R.}~\bibnamefont {Kleiner}},\ }\href
  {\doibase 10.1038/nphoton.2013.216} {\bibfield  {journal} {\bibinfo
  {journal} {Nat Photon}\ }\textbf {\bibinfo {volume} {7}},\ \bibinfo {pages}
  {702} (\bibinfo {year} {2013})}\BibitemShut {NoStop}%
\bibitem [{\citenamefont {Cariglia}\ \emph {et~al.}(2014)\citenamefont
  {Cariglia}, \citenamefont {Vargas-Paredes},\ and\ \citenamefont
  {Doria}}]{cariglia14}%
  \BibitemOpen
  \bibfield  {author} {\bibinfo {author} {\bibfnamefont {M.}~\bibnamefont
  {Cariglia}}, \bibinfo {author} {\bibfnamefont {A.~A.}\ \bibnamefont
  {Vargas-Paredes}}, \ and\ \bibinfo {author} {\bibfnamefont {M.~M.}\
  \bibnamefont {Doria}},\ }\href
  {http://stacks.iop.org/0295-5075/105/i=3/a=31002} {\bibfield  {journal}
  {\bibinfo  {journal} {Europhysics Letters}\ }\textbf {\bibinfo {volume}
  {105}},\ \bibinfo {pages} {31002} (\bibinfo {year} {2014})}\BibitemShut
  {NoStop}%
\bibitem [{\citenamefont {Berg}\ \emph {et~al.}(2009)\citenamefont {Berg},
  \citenamefont {Fradkin}, \citenamefont {Kivelson},\ and\ \citenamefont
  {Tranquada}}]{berg09}%
  \BibitemOpen
  \bibfield  {author} {\bibinfo {author} {\bibfnamefont {E.}~\bibnamefont
  {Berg}}, \bibinfo {author} {\bibfnamefont {E.}~\bibnamefont {Fradkin}},
  \bibinfo {author} {\bibfnamefont {S.~A.}\ \bibnamefont {Kivelson}}, \ and\
  \bibinfo {author} {\bibfnamefont {J.~M.}\ \bibnamefont {Tranquada}},\ }\href
  {http://stacks.iop.org/1367-2630/11/i=11/a=115004} {\bibfield  {journal}
  {\bibinfo  {journal} {New Journal of Physics}\ }\textbf {\bibinfo {volume}
  {11}},\ \bibinfo {pages} {115004} (\bibinfo {year} {2009})}\BibitemShut
  {NoStop}%
\bibitem [{\citenamefont {Vojta}(2009)}]{vojta09}%
  \BibitemOpen
  \bibfield  {author} {\bibinfo {author} {\bibfnamefont {M.}~\bibnamefont
  {Vojta}},\ }\href {\doibase 10.1080/00018730903122242} {\bibfield  {journal}
  {\bibinfo  {journal} {Advances in Physics}\ }\textbf {\bibinfo {volume}
  {58}},\ \bibinfo {pages} {699} (\bibinfo {year} {2009})}\BibitemShut
  {NoStop}%
\bibitem [{\citenamefont {Khasanov}\ \emph {et~al.}(2007)\citenamefont
  {Khasanov}, \citenamefont {Shengelaya}, \citenamefont {Maisuradze},
  \citenamefont {Mattina}, \citenamefont {Bussmann-Holder}, \citenamefont
  {Keller},\ and\ \citenamefont {M\"uller}}]{khasanov07}%
  \BibitemOpen
  \bibfield  {author} {\bibinfo {author} {\bibfnamefont {R.}~\bibnamefont
  {Khasanov}}, \bibinfo {author} {\bibfnamefont {A.}~\bibnamefont
  {Shengelaya}}, \bibinfo {author} {\bibfnamefont {A.}~\bibnamefont
  {Maisuradze}}, \bibinfo {author} {\bibfnamefont {F.~L.}\ \bibnamefont
  {Mattina}}, \bibinfo {author} {\bibfnamefont {A.}~\bibnamefont
  {Bussmann-Holder}}, \bibinfo {author} {\bibfnamefont {H.}~\bibnamefont
  {Keller}}, \ and\ \bibinfo {author} {\bibfnamefont {K.~A.}\ \bibnamefont
  {M\"uller}},\ }\href {\doibase 10.1103/PhysRevLett.98.057007} {\bibfield
  {journal} {\bibinfo  {journal} {Phys. Rev. Lett.}\ }\textbf {\bibinfo
  {volume} {98}},\ \bibinfo {pages} {057007} (\bibinfo {year}
  {2007})}\BibitemShut {NoStop}%
\bibitem [{\citenamefont {Alloul}\ \emph {et~al.}(1989)\citenamefont {Alloul},
  \citenamefont {Ohno},\ and\ \citenamefont {Mendels}}]{alloul89}%
  \BibitemOpen
  \bibfield  {author} {\bibinfo {author} {\bibfnamefont {H.}~\bibnamefont
  {Alloul}}, \bibinfo {author} {\bibfnamefont {T.}~\bibnamefont {Ohno}}, \ and\
  \bibinfo {author} {\bibfnamefont {P.}~\bibnamefont {Mendels}},\ }\href
  {\doibase 10.1103/PhysRevLett.63.1700} {\bibfield  {journal} {\bibinfo
  {journal} {Phys. Rev. Lett.}\ }\textbf {\bibinfo {volume} {63}},\ \bibinfo
  {pages} {1700} (\bibinfo {year} {1989})}\BibitemShut {NoStop}%
\bibitem [{\citenamefont {He}\ \emph {et~al.}(2011)\citenamefont {He},
  \citenamefont {Hashimoto}, \citenamefont {Karapetyan}, \citenamefont
  {Koralek}, \citenamefont {Hinton}, \citenamefont {Testaud}, \citenamefont
  {Nathan}, \citenamefont {Yoshida}, \citenamefont {Yao}, \citenamefont
  {Tanaka}, \citenamefont {Meevasana}, \citenamefont {Moore}, \citenamefont
  {Lu}, \citenamefont {Mo}, \citenamefont {Ishikado}, \citenamefont {Eisaki},
  \citenamefont {Hussain}, \citenamefont {Devereaux}, \citenamefont {Kivelson},
  \citenamefont {Orenstein}, \citenamefont {Kapitulnik},\ and\ \citenamefont
  {Shen}}]{he11}%
  \BibitemOpen
  \bibfield  {author} {\bibinfo {author} {\bibfnamefont {R.-H.}\ \bibnamefont
  {He}}, \bibinfo {author} {\bibfnamefont {M.}~\bibnamefont {Hashimoto}},
  \bibinfo {author} {\bibfnamefont {H.}~\bibnamefont {Karapetyan}}, \bibinfo
  {author} {\bibfnamefont {J.~D.}\ \bibnamefont {Koralek}}, \bibinfo {author}
  {\bibfnamefont {J.~P.}\ \bibnamefont {Hinton}}, \bibinfo {author}
  {\bibfnamefont {J.~P.}\ \bibnamefont {Testaud}}, \bibinfo {author}
  {\bibfnamefont {V.}~\bibnamefont {Nathan}}, \bibinfo {author} {\bibfnamefont
  {Y.}~\bibnamefont {Yoshida}}, \bibinfo {author} {\bibfnamefont
  {H.}~\bibnamefont {Yao}}, \bibinfo {author} {\bibfnamefont {K.}~\bibnamefont
  {Tanaka}}, \bibinfo {author} {\bibfnamefont {W.}~\bibnamefont {Meevasana}},
  \bibinfo {author} {\bibfnamefont {R.~G.}\ \bibnamefont {Moore}}, \bibinfo
  {author} {\bibfnamefont {D.~H.}\ \bibnamefont {Lu}}, \bibinfo {author}
  {\bibfnamefont {S.-K.}\ \bibnamefont {Mo}}, \bibinfo {author} {\bibfnamefont
  {M.}~\bibnamefont {Ishikado}}, \bibinfo {author} {\bibfnamefont
  {H.}~\bibnamefont {Eisaki}}, \bibinfo {author} {\bibfnamefont
  {Z.}~\bibnamefont {Hussain}}, \bibinfo {author} {\bibfnamefont {T.~P.}\
  \bibnamefont {Devereaux}}, \bibinfo {author} {\bibfnamefont {S.~A.}\
  \bibnamefont {Kivelson}}, \bibinfo {author} {\bibfnamefont {J.}~\bibnamefont
  {Orenstein}}, \bibinfo {author} {\bibfnamefont {A.}~\bibnamefont
  {Kapitulnik}}, \ and\ \bibinfo {author} {\bibfnamefont {Z.-X.}\ \bibnamefont
  {Shen}},\ }\href {\doibase 10.1126/science.1198415} {\bibfield  {journal}
  {\bibinfo  {journal} {Science}\ }\textbf {\bibinfo {volume} {331}},\ \bibinfo
  {pages} {1579} (\bibinfo {year} {2011})}\BibitemShut {NoStop}%
\bibitem [{\citenamefont {Daou}\ \emph {et~al.}(2010)\citenamefont {Daou},
  \citenamefont {Chang}, \citenamefont {LeBoeuf}, \citenamefont
  {Cyr-Choiniere}, \citenamefont {Laliberte}, \citenamefont {Doiron-Leyraud},
  \citenamefont {Ramshaw}, \citenamefont {Liang}, \citenamefont {Bonn},
  \citenamefont {Hardy},\ and\ \citenamefont {Taillefer}}]{daou10}%
  \BibitemOpen
  \bibfield  {author} {\bibinfo {author} {\bibfnamefont {R.}~\bibnamefont
  {Daou}}, \bibinfo {author} {\bibfnamefont {J.}~\bibnamefont {Chang}},
  \bibinfo {author} {\bibfnamefont {D.}~\bibnamefont {LeBoeuf}}, \bibinfo
  {author} {\bibfnamefont {O.}~\bibnamefont {Cyr-Choiniere}}, \bibinfo {author}
  {\bibfnamefont {F.}~\bibnamefont {Laliberte}}, \bibinfo {author}
  {\bibfnamefont {N.}~\bibnamefont {Doiron-Leyraud}}, \bibinfo {author}
  {\bibfnamefont {B.~J.}\ \bibnamefont {Ramshaw}}, \bibinfo {author}
  {\bibfnamefont {R.}~\bibnamefont {Liang}}, \bibinfo {author} {\bibfnamefont
  {D.~A.}\ \bibnamefont {Bonn}}, \bibinfo {author} {\bibfnamefont {W.~N.}\
  \bibnamefont {Hardy}}, \ and\ \bibinfo {author} {\bibfnamefont
  {L.}~\bibnamefont {Taillefer}},\ }\href {\doibase 10.1038/nature08716}
  {\bibfield  {journal} {\bibinfo  {journal} {Nature}\ }\textbf {\bibinfo
  {volume} {463}},\ \bibinfo {pages} {519} (\bibinfo {year}
  {2010})}\BibitemShut {NoStop}%
\bibitem [{\citenamefont {Volovik}\ and\ \citenamefont
  {Gor'kov}(1985)}]{volovik85}%
  \BibitemOpen
  \bibfield  {author} {\bibinfo {author} {\bibfnamefont {G.~E.}\ \bibnamefont
  {Volovik}}\ and\ \bibinfo {author} {\bibfnamefont {L.~P.}\ \bibnamefont
  {Gor'kov}},\ }\href@noop {} {\bibfield  {journal} {\bibinfo  {journal} {Sov.
  Phys. JETP}\ }\textbf {\bibinfo {volume} {61}},\ \bibinfo {pages} {843}
  (\bibinfo {year} {1985})}\BibitemShut {NoStop}%
\bibitem [{\citenamefont {Abrikosov}(1957)}]{abrikosov57}%
  \BibitemOpen
  \bibfield  {author} {\bibinfo {author} {\bibfnamefont {A.~A.}\ \bibnamefont
  {Abrikosov}},\ }\href@noop {} {\bibfield  {journal} {\bibinfo  {journal}
  {Soviet Physics JETP}\ }\textbf {\bibinfo {volume} {5}},\ \bibinfo {pages}
  {1174} (\bibinfo {year} {1957})}\BibitemShut {NoStop}%
\bibitem [{\citenamefont {Bogomolny}(1976)}]{bogomolny76}%
  \BibitemOpen
  \bibfield  {author} {\bibinfo {author} {\bibfnamefont {E.~B.}\ \bibnamefont
  {Bogomolny}},\ }\href@noop {} {\bibfield  {journal} {\bibinfo  {journal}
  {Sov. J. Nucl. Phys.}\ }\textbf {\bibinfo {volume} {24}},\ \bibinfo {pages}
  {449} (\bibinfo {year} {1976})}\BibitemShut {NoStop}%
\bibitem [{\citenamefont {Seiberg}\ and\ \citenamefont
  {Witten}(1994)}]{seiberg94}%
  \BibitemOpen
  \bibfield  {author} {\bibinfo {author} {\bibfnamefont {N.}~\bibnamefont
  {Seiberg}}\ and\ \bibinfo {author} {\bibfnamefont {E.}~\bibnamefont
  {Witten}},\ }\href {\doibase 10.1016/0550-3213(94)90124-4} {\bibfield
  {journal} {\bibinfo  {journal} {Nuclear Physics B}\ }\textbf {\bibinfo
  {volume} {426}},\ \bibinfo {pages} {19 } (\bibinfo {year}
  {1994})}\BibitemShut {NoStop}%
\bibitem [{\citenamefont {Doria}\ \emph {et~al.}(2010)\citenamefont {Doria},
  \citenamefont {de~C.~Romaguera},\ and\ \citenamefont {Peeters}}]{doria10}%
  \BibitemOpen
  \bibfield  {author} {\bibinfo {author} {\bibfnamefont {M.~M.}\ \bibnamefont
  {Doria}}, \bibinfo {author} {\bibfnamefont {A.~R.}\ \bibnamefont
  {de~C.~Romaguera}}, \ and\ \bibinfo {author} {\bibfnamefont {F.~M.}\
  \bibnamefont {Peeters}},\ }\href@noop {} {\bibfield  {journal} {\bibinfo
  {journal} {Europhys. Lett.}\ }\textbf {\bibinfo {volume} {92}},\ \bibinfo
  {pages} {17004} (\bibinfo {year} {2010})}\BibitemShut {NoStop}%
\bibitem [{\citenamefont {Vargas-Paredes}\ \emph
  {et~al.}(2014{\natexlab{a}})\citenamefont {Vargas-Paredes}, \citenamefont
  {Cariglia}, \citenamefont {Doria}, \citenamefont {Rodrigues},\ and\
  \citenamefont {C.~Romaguera}}]{edinardo14}%
  \BibitemOpen
  \bibfield  {author} {\bibinfo {author} {\bibfnamefont {A.~A.}\ \bibnamefont
  {Vargas-Paredes}}, \bibinfo {author} {\bibfnamefont {M.}~\bibnamefont
  {Cariglia}}, \bibinfo {author} {\bibfnamefont {M.~M.}\ \bibnamefont {Doria}},
  \bibinfo {author} {\bibfnamefont {E.~I.}\ \bibnamefont {Rodrigues}}, \ and\
  \bibinfo {author} {\bibfnamefont {A.}~\bibnamefont {C.~Romaguera}},\ }\href
  {\doibase 10.1007/s10948-013-2310-5} {\bibfield  {journal} {\bibinfo
  {journal} {Journal of Superconductivity and Novel Magnetism}\ }\textbf
  {\bibinfo {volume} {27}},\ \bibinfo {pages} {349} (\bibinfo {year}
  {2014}{\natexlab{a}})}\BibitemShut {NoStop}%
\bibitem [{\citenamefont {Rai\ifmmode \check{c}\else
  \v{c}\fi{}evi\ifmmode~\acute{c}\else \'{c}\fi{}}\ \emph
  {et~al.}(2011)\citenamefont {Rai\ifmmode \check{c}\else
  \v{c}\fi{}evi\ifmmode~\acute{c}\else \'{c}\fi{}}, \citenamefont
  {Popovi\ifmmode~\acute{c}\else \'{c}\fi{}}, \citenamefont {Panagopoulos},
  \citenamefont {Benfatto}, \citenamefont {Silva~Neto}, \citenamefont {Choi},\
  and\ \citenamefont {Sasagawa}}]{marcelo11}%
  \BibitemOpen
  \bibfield  {author} {\bibinfo {author} {\bibfnamefont {I.}~\bibnamefont
  {Rai\ifmmode \check{c}\else \v{c}\fi{}evi\ifmmode~\acute{c}\else
  \'{c}\fi{}}}, \bibinfo {author} {\bibfnamefont {D.}~\bibnamefont
  {Popovi\ifmmode~\acute{c}\else \'{c}\fi{}}}, \bibinfo {author} {\bibfnamefont
  {C.}~\bibnamefont {Panagopoulos}}, \bibinfo {author} {\bibfnamefont
  {L.}~\bibnamefont {Benfatto}}, \bibinfo {author} {\bibfnamefont {M.~B.}\
  \bibnamefont {Silva~Neto}}, \bibinfo {author} {\bibfnamefont {E.~S.}\
  \bibnamefont {Choi}}, \ and\ \bibinfo {author} {\bibfnamefont
  {T.}~\bibnamefont {Sasagawa}},\ }\href {\doibase
  10.1103/PhysRevLett.106.227206} {\bibfield  {journal} {\bibinfo  {journal}
  {Phys. Rev. Lett.}\ }\textbf {\bibinfo {volume} {106}},\ \bibinfo {pages}
  {227206} (\bibinfo {year} {2011})}\BibitemShut {NoStop}%
\bibitem [{\citenamefont {Yanagisawa}\ \emph {et~al.}(2012)\citenamefont
  {Yanagisawa}, \citenamefont {Tanaka}, \citenamefont {Hase},\ and\
  \citenamefont {Yamaji}}]{takashi12}%
  \BibitemOpen
  \bibfield  {author} {\bibinfo {author} {\bibfnamefont {T.}~\bibnamefont
  {Yanagisawa}}, \bibinfo {author} {\bibfnamefont {Y.}~\bibnamefont {Tanaka}},
  \bibinfo {author} {\bibfnamefont {I.}~\bibnamefont {Hase}}, \ and\ \bibinfo
  {author} {\bibfnamefont {K.}~\bibnamefont {Yamaji}},\ }\href {\doibase
  10.1143/JPSJ.81.024712} {\bibfield  {journal} {\bibinfo  {journal} {Journal
  of the Physical Society of Japan}\ }\textbf {\bibinfo {volume} {81}},\
  \bibinfo {pages} {024712} (\bibinfo {year} {2012})}\BibitemShut {NoStop}%
\bibitem [{\citenamefont {Kaminski}\ \emph {et~al.}(2002)\citenamefont
  {Kaminski}, \citenamefont {Fretwell}, \citenamefont {Campuzano},
  \citenamefont {Li}, \citenamefont {Raffy}, \citenamefont {Cullen},
  \citenamefont {You}, \citenamefont {Olson}, \citenamefont {Varma},\ and\
  \citenamefont {Hochst}}]{kaminski02}%
  \BibitemOpen
  \bibfield  {author} {\bibinfo {author} {\bibfnamefont {S.}~\bibnamefont
  {Kaminski}, \bibfnamefont {A.~Rosenkranz}}, \bibinfo {author} {\bibfnamefont
  {H.~M.}\ \bibnamefont {Fretwell}}, \bibinfo {author} {\bibfnamefont {J.~C.}\
  \bibnamefont {Campuzano}}, \bibinfo {author} {\bibfnamefont {Z.}~\bibnamefont
  {Li}}, \bibinfo {author} {\bibfnamefont {H.}~\bibnamefont {Raffy}}, \bibinfo
  {author} {\bibfnamefont {W.~G.}\ \bibnamefont {Cullen}}, \bibinfo {author}
  {\bibfnamefont {H.}~\bibnamefont {You}}, \bibinfo {author} {\bibfnamefont
  {C.~G.}\ \bibnamefont {Olson}}, \bibinfo {author} {\bibfnamefont {C.~M.}\
  \bibnamefont {Varma}}, \ and\ \bibinfo {author} {\bibfnamefont
  {H.}~\bibnamefont {Hochst}},\ }\href {\doibase
  http://dx.doi.org/10.1038/416610a} {\bibfield  {journal} {\bibinfo  {journal}
  {Nature}\ }\textbf {\bibinfo {volume} {416}},\ \bibinfo {pages} {610}
  (\bibinfo {year} {2002})}\BibitemShut {NoStop}%
\bibitem [{\citenamefont {Str\"assle}\ \emph {et~al.}(2008)\citenamefont
  {Str\"assle}, \citenamefont {Roos}, \citenamefont {Mali}, \citenamefont
  {Keller},\ and\ \citenamefont {Ohno}}]{strassle08}%
  \BibitemOpen
  \bibfield  {author} {\bibinfo {author} {\bibfnamefont {S.}~\bibnamefont
  {Str\"assle}}, \bibinfo {author} {\bibfnamefont {J.}~\bibnamefont {Roos}},
  \bibinfo {author} {\bibfnamefont {M.}~\bibnamefont {Mali}}, \bibinfo {author}
  {\bibfnamefont {H.}~\bibnamefont {Keller}}, \ and\ \bibinfo {author}
  {\bibfnamefont {T.}~\bibnamefont {Ohno}},\ }\href {\doibase
  10.1103/PhysRevLett.101.237001} {\bibfield  {journal} {\bibinfo  {journal}
  {Phys. Rev. Lett.}\ }\textbf {\bibinfo {volume} {101}},\ \bibinfo {pages}
  {237001} (\bibinfo {year} {2008})}\BibitemShut {NoStop}%
\bibitem [{\citenamefont {Str\"assle}\ \emph {et~al.}(2011)\citenamefont
  {Str\"assle}, \citenamefont {Graneli}, \citenamefont {Mali}, \citenamefont
  {Roos},\ and\ \citenamefont {Keller}}]{strassle11}%
  \BibitemOpen
  \bibfield  {author} {\bibinfo {author} {\bibfnamefont {S.}~\bibnamefont
  {Str\"assle}}, \bibinfo {author} {\bibfnamefont {B.}~\bibnamefont {Graneli}},
  \bibinfo {author} {\bibfnamefont {M.}~\bibnamefont {Mali}}, \bibinfo {author}
  {\bibfnamefont {J.}~\bibnamefont {Roos}}, \ and\ \bibinfo {author}
  {\bibfnamefont {H.}~\bibnamefont {Keller}},\ }\href {\doibase
  10.1103/PhysRevLett.106.097003} {\bibfield  {journal} {\bibinfo  {journal}
  {Phys. Rev. Lett.}\ }\textbf {\bibinfo {volume} {106}},\ \bibinfo {pages}
  {097003} (\bibinfo {year} {2011})}\BibitemShut {NoStop}%
\bibitem [{\citenamefont {MacDougall}\ \emph {et~al.}(2008)\citenamefont
  {MacDougall}, \citenamefont {Aczel}, \citenamefont {Carlo}, \citenamefont
  {Ito}, \citenamefont {Rodriguez}, \citenamefont {Russo}, \citenamefont
  {Uemura}, \citenamefont {Wakimoto},\ and\ \citenamefont
  {Luke}}]{macdougall08}%
  \BibitemOpen
  \bibfield  {author} {\bibinfo {author} {\bibfnamefont {G.~J.}\ \bibnamefont
  {MacDougall}}, \bibinfo {author} {\bibfnamefont {A.~A.}\ \bibnamefont
  {Aczel}}, \bibinfo {author} {\bibfnamefont {J.~P.}\ \bibnamefont {Carlo}},
  \bibinfo {author} {\bibfnamefont {T.}~\bibnamefont {Ito}}, \bibinfo {author}
  {\bibfnamefont {J.}~\bibnamefont {Rodriguez}}, \bibinfo {author}
  {\bibfnamefont {P.~L.}\ \bibnamefont {Russo}}, \bibinfo {author}
  {\bibfnamefont {Y.~J.}\ \bibnamefont {Uemura}}, \bibinfo {author}
  {\bibfnamefont {S.}~\bibnamefont {Wakimoto}}, \ and\ \bibinfo {author}
  {\bibfnamefont {G.~M.}\ \bibnamefont {Luke}},\ }\href {\doibase
  10.1103/PhysRevLett.101.017001} {\bibfield  {journal} {\bibinfo  {journal}
  {Phys. Rev. Lett.}\ }\textbf {\bibinfo {volume} {101}},\ \bibinfo {pages}
  {017001} (\bibinfo {year} {2008})}\BibitemShut {NoStop}%
\bibitem [{\citenamefont {Sonier}\ \emph {et~al.}(2009)\citenamefont {Sonier},
  \citenamefont {Pacradouni}, \citenamefont {Sabok-Sayr}, \citenamefont
  {Hardy}, \citenamefont {Bonn}, \citenamefont {Liang},\ and\ \citenamefont
  {Mook}}]{sonier09}%
  \BibitemOpen
  \bibfield  {author} {\bibinfo {author} {\bibfnamefont {J.~E.}\ \bibnamefont
  {Sonier}}, \bibinfo {author} {\bibfnamefont {V.}~\bibnamefont {Pacradouni}},
  \bibinfo {author} {\bibfnamefont {S.~A.}\ \bibnamefont {Sabok-Sayr}},
  \bibinfo {author} {\bibfnamefont {W.~N.}\ \bibnamefont {Hardy}}, \bibinfo
  {author} {\bibfnamefont {D.~A.}\ \bibnamefont {Bonn}}, \bibinfo {author}
  {\bibfnamefont {R.}~\bibnamefont {Liang}}, \ and\ \bibinfo {author}
  {\bibfnamefont {H.~A.}\ \bibnamefont {Mook}},\ }\href {\doibase
  10.1103/PhysRevLett.103.167002} {\bibfield  {journal} {\bibinfo  {journal}
  {Phys. Rev. Lett.}\ }\textbf {\bibinfo {volume} {103}},\ \bibinfo {pages}
  {167002} (\bibinfo {year} {2009})}\BibitemShut {NoStop}%
\bibitem [{\citenamefont {Vargas-Paredes}\ \emph {et~al.}(2013)\citenamefont
  {Vargas-Paredes}, \citenamefont {Doria},\ and\ \citenamefont
  {Neto}}]{alfredo13}%
  \BibitemOpen
  \bibfield  {author} {\bibinfo {author} {\bibfnamefont {A.~A.}\ \bibnamefont
  {Vargas-Paredes}}, \bibinfo {author} {\bibfnamefont {M.~M.}\ \bibnamefont
  {Doria}}, \ and\ \bibinfo {author} {\bibfnamefont {J.~A.~H.}\ \bibnamefont
  {Neto}},\ }\href {\doibase 10.1063/1.4773286} {\bibfield  {journal} {\bibinfo
   {journal} {Journal of Mathematical Physics}\ }\textbf {\bibinfo {volume}
  {54}},\ \bibinfo {eid} {013101} (\bibinfo {year} {2013})}\BibitemShut
  {NoStop}%
\bibitem [{\citenamefont {Vargas-Paredes}\ \emph
  {et~al.}(2014{\natexlab{b}})\citenamefont {Vargas-Paredes}, \citenamefont
  {Doria},\ and\ \citenamefont {Cariglia}}]{alfredo14}%
  \BibitemOpen
  \bibfield  {author} {\bibinfo {author} {\bibfnamefont {A.~A.}\ \bibnamefont
  {Vargas-Paredes}}, \bibinfo {author} {\bibfnamefont {M.~M.}\ \bibnamefont
  {Doria}}, \ and\ \bibinfo {author} {\bibfnamefont {M.}~\bibnamefont
  {Cariglia}},\ }\href@noop {} {\bibfield  {journal} {\bibinfo  {journal} {\it
  arXiv:1407.4030}\ } (\bibinfo {year} {2014}{\natexlab{b}})}\BibitemShut
  {NoStop}%
\bibitem [{\citenamefont {Islam}\ \emph {et~al.}(2004)\citenamefont {Islam},
  \citenamefont {Liu}, \citenamefont {Sinha}, \citenamefont {Lang},
  \citenamefont {Moss}, \citenamefont {Haskel}, \citenamefont {Srajer},
  \citenamefont {Wochner}, \citenamefont {Lee}, \citenamefont {Haeffner},\ and\
  \citenamefont {Welp}}]{zahirul04}%
  \BibitemOpen
  \bibfield  {author} {\bibinfo {author} {\bibfnamefont {Z.}~\bibnamefont
  {Islam}}, \bibinfo {author} {\bibfnamefont {X.}~\bibnamefont {Liu}}, \bibinfo
  {author} {\bibfnamefont {S.~K.}\ \bibnamefont {Sinha}}, \bibinfo {author}
  {\bibfnamefont {J.~C.}\ \bibnamefont {Lang}}, \bibinfo {author}
  {\bibfnamefont {S.~C.}\ \bibnamefont {Moss}}, \bibinfo {author}
  {\bibfnamefont {D.}~\bibnamefont {Haskel}}, \bibinfo {author} {\bibfnamefont
  {G.}~\bibnamefont {Srajer}}, \bibinfo {author} {\bibfnamefont
  {P.}~\bibnamefont {Wochner}}, \bibinfo {author} {\bibfnamefont {D.~R.}\
  \bibnamefont {Lee}}, \bibinfo {author} {\bibfnamefont {D.~R.}\ \bibnamefont
  {Haeffner}}, \ and\ \bibinfo {author} {\bibfnamefont {U.}~\bibnamefont
  {Welp}},\ }\href {\doibase 10.1103/PhysRevLett.93.157008} {\bibfield
  {journal} {\bibinfo  {journal} {Phys. Rev. Lett.}\ }\textbf {\bibinfo
  {volume} {93}},\ \bibinfo {pages} {157008} (\bibinfo {year}
  {2004})}\BibitemShut {NoStop}%
\bibitem [{\citenamefont {Ghiringhelli}\ \emph {et~al.}(2012)\citenamefont
  {Ghiringhelli}, \citenamefont {Le~Tacon}, \citenamefont {Minola},
  \citenamefont {Blanco-Canosa}, \citenamefont {Mazzoli}, \citenamefont
  {Brookes}, \citenamefont {De~Luca}, \citenamefont {Frano}, \citenamefont
  {Hawthorn}, \citenamefont {He}, \citenamefont {Loew}, \citenamefont {Sala},
  \citenamefont {Peets}, \citenamefont {Salluzzo}, \citenamefont {Schierle},
  \citenamefont {Sutarto}, \citenamefont {Sawatzky}, \citenamefont {Weschke},
  \citenamefont {Keimer},\ and\ \citenamefont {Braicovich}}]{ghiringhelli12}%
  \BibitemOpen
  \bibfield  {author} {\bibinfo {author} {\bibfnamefont {G.}~\bibnamefont
  {Ghiringhelli}}, \bibinfo {author} {\bibfnamefont {M.}~\bibnamefont
  {Le~Tacon}}, \bibinfo {author} {\bibfnamefont {M.}~\bibnamefont {Minola}},
  \bibinfo {author} {\bibfnamefont {S.}~\bibnamefont {Blanco-Canosa}}, \bibinfo
  {author} {\bibfnamefont {C.}~\bibnamefont {Mazzoli}}, \bibinfo {author}
  {\bibfnamefont {N.~B.}\ \bibnamefont {Brookes}}, \bibinfo {author}
  {\bibfnamefont {G.~M.}\ \bibnamefont {De~Luca}}, \bibinfo {author}
  {\bibfnamefont {A.}~\bibnamefont {Frano}}, \bibinfo {author} {\bibfnamefont
  {D.~G.}\ \bibnamefont {Hawthorn}}, \bibinfo {author} {\bibfnamefont
  {F.}~\bibnamefont {He}}, \bibinfo {author} {\bibfnamefont {T.}~\bibnamefont
  {Loew}}, \bibinfo {author} {\bibfnamefont {M.~M.}\ \bibnamefont {Sala}},
  \bibinfo {author} {\bibfnamefont {D.~C.}\ \bibnamefont {Peets}}, \bibinfo
  {author} {\bibfnamefont {M.}~\bibnamefont {Salluzzo}}, \bibinfo {author}
  {\bibfnamefont {E.}~\bibnamefont {Schierle}}, \bibinfo {author}
  {\bibfnamefont {R.}~\bibnamefont {Sutarto}}, \bibinfo {author} {\bibfnamefont
  {G.~A.}\ \bibnamefont {Sawatzky}}, \bibinfo {author} {\bibfnamefont
  {E.}~\bibnamefont {Weschke}}, \bibinfo {author} {\bibfnamefont
  {B.}~\bibnamefont {Keimer}}, \ and\ \bibinfo {author} {\bibfnamefont
  {L.}~\bibnamefont {Braicovich}},\ }\href {\doibase 10.1126/science.1223532}
  {\ \textbf {\bibinfo {volume} {337}},\ \bibinfo {pages} {821} (\bibinfo
  {year} {2012})}\BibitemShut {NoStop}%
\end{thebibliography}%
\end{document}